\documentclass[twocolumn,fleqn,usenatbib,linenumber]{aastex63}

\usepackage{mathptmx}
\usepackage[T1]{fontenc}
\usepackage{ae,aecompl}

\usepackage{graphicx}   
\usepackage{amsmath}    
\usepackage{amssymb}   

\usepackage{lineno}

\usepackage{xcolor} 

\received{XX XXXX}
\revised{XX XXXX}
\accepted{XX XXXX}
\submitjournal{ApJ}

\shorttitle{Nova Rate}
\shortauthors{Kawash et al.}

\begin{document}

\title{The Galactic Nova Rate: Estimates from the ASAS-SN and \textit{Gaia} Surveys}

\correspondingauthor{Adam Kawash}
\email{kawashad@msu.edu}

\author[0000-0003-0071-1622]{A.\ Kawash}
\affiliation{Center for Data Intensive and Time Domain Astronomy, Department of Physics and Astronomy, Michigan State University, East Lansing, MI 48824, USA}

\author[0000-0002-8400-3705]{L.\ Chomiuk}
\affiliation{Center for Data Intensive and Time Domain Astronomy, Department of Physics and Astronomy, Michigan State University, East Lansing, MI 48824, USA}

\author[0000-0002-1468-9668]{J.\ Strader}
\affiliation{Center for Data Intensive and Time Domain Astronomy, Department of Physics and Astronomy, Michigan State University, East Lansing, MI 48824, USA}

\author[0000-0001-5991-6863]{K.~V.\ Sokolovsky}
\affiliation{Center for Data Intensive and Time Domain Astronomy, Department of Physics and Astronomy, Michigan State University, East Lansing, MI 48824, USA}
\affiliation{Sternberg Astronomical Institute, Moscow State University, Universitetskii~pr.~13, 119992~Moscow, Russia}

\author[0000-0001-8525-3442]{E.\ Aydi}
\affiliation{Center for Data Intensive and Time Domain Astronomy, Department of Physics and Astronomy, Michigan State University, East Lansing, MI 48824, USA}

\author{C. S. \ Kochanek}
\affiliation{Department of Astronomy, The Ohio State University, 140 West 18th Avenue, Columbus, OH 43210, USA}
\affiliation{Center for Cosmology and Astroparticle Physics, The Ohio State University, 191 W. Woodruff Avenue, Columbus, OH 43210, USA}

\author{K.\ Z.\ Stanek}
\affiliation{Department of Astronomy, The Ohio State University, 140 West 18th Avenue, Columbus, OH 43210, USA}
\affiliation{Center for Cosmology and Astroparticle Physics, The Ohio State University, 191 W. Woodruff Avenue, Columbus, OH 43210, USA}

\author{Z.\ Kostrzewa-Rutkowska}
\affiliation{Leiden Observatory, Leiden University, PO Box 9513, 2300 RA
Leiden, The Netherlands}
\affiliation{SRON Netherlands Institute for Space Research, Niels Bohrweg 4, 2333 CA Leiden, The Netherlands}

\author{S.~T.\ Hodgkin}
\affiliation{Institute of Astronomy, Madingley Road, Cambridge CB3 0HA,
UK.}

\author[0000-0002-8286-8094]{K.\ Mukai}
\affiliation{CRESST II and X-ray Astrophysics Laboratory, NASA/GSFC, Greenbelt, MD 20771, USA}
\affiliation{Department of Physics, University of Maryland, Baltimore County, 1000 Hilltop Circle, Baltimore, MD 21250, USA}

\author{B.\ Shappee}
\affiliation{Institute for Astronomy, University of Hawai`i at M\=anoa, 2680 Woodlawn Dr., Honolulu 96822, USA}

\author{T.\ Jayasinghe}
\affiliation{Department of Astronomy, The Ohio State University, 140 West 18th Avenue, Columbus, OH 43210, USA}

\author{M.\ Rizzo Smith}
\affiliation{Department of Astronomy, The Ohio State University, 140 West 18th Avenue, Columbus, OH 43210, USA}

\author{ T.~W.-S.\ Holoien}
\affiliation{Carnegie Observatories, 813 Santa Barbara Street, Pasadena, CA 91101, USA}

\author{J.~L.\ Prieto}
\affiliation{N\'ucleo de Astronom\'ia de la Facultad de Ingenier\'ia y Ciencias, Universidad Diego Portales, Av. Ej\'ercito 441, Santiago, Chile}
\affiliation{Millennium Institute of Astrophysics, Santiago, Chile }

\author{T.~A.\ Thompson}
\affiliation{Department of Astronomy, The Ohio State University, 140 West 18th Avenue, Columbus, OH 43210, USA}
\affiliation{Center for Cosmology and Astroparticle Physics, The Ohio State University, 191 W. Woodruff Avenue, Columbus, OH 43210, USA}

\begin{abstract}
We present the first estimate of the Galactic nova rate based on optical transient surveys covering the entire sky. Using data from the All-Sky Automated Survey for Supernovae (ASAS-SN) and \textit{Gaia}---the only two all-sky surveys to report classical nova candidates---we find 39 confirmed Galactic novae and 7 additional unconfirmed candidates discovered from 2019--2021, yielding a nova discovery rate of $\approx 14$ yr$^{-1}$. Using accurate Galactic stellar mass models,  three-dimensional dust maps, and incorporating realistic nova light curves, we have built a sophisticated Galactic nova model that allows an estimate of the recovery fraction of Galactic novae from these surveys over this time period. The observing capabilities of each survey are distinct: the high cadence of ASAS-SN makes it sensitive to fast novae, while the broad observing filter and high spatial resolution of \textit{Gaia} make it more sensitive to highly reddened novae across the entire Galactic plane and bulge. Despite these differences, we find that ASAS-SN and \textit{Gaia} give consistent Galactic nova rates, with a final joint nova rate of $26 \pm 5$ yr$^{-1}$. This inferred nova rate is substantially lower than found by many other recent studies. Critically assessing the systematic uncertainties in the Galactic nova rate, we argue that the role of faint fast-fading novae has likely been overestimated, but that subtle details in the operation of transient alert pipelines can have large, sometimes unappreciated effects on transient recovery efficiency. Our predicted nova rate can be directly tested with forthcoming red/near-infrared transient surveys in the southern hemisphere.
\end{abstract}


\section{Introduction} \label{sec:intro}

A classical nova eruption is the result of a thermonuclear runaway of accreted hydrogen rich material on the surface of a white dwarf (see \citealt{be08,Chomiuk21} for reviews). 
At peak brightness, novae are relatively luminous, with absolute magnitudes between M$_V$ $\approx$ $-4$ to $-10$ mag \citep{shafter_2017}, allowing them to be discovered out to the largest Galactic distances and 
in nearby galaxies. Discoveries of Galactic novae date back thousands of years \citep{patterson13,shara17}, and estimates of the total frequency with which the Milky Way produces novae date back nearly a century \citep{lund35L}. The nova rate has broad implications in a range of areas, including Galactic nucleosyntheis, binary evolution, and the origin of Type Ia supernovae.

Found in both the Galactic disk and bulge, novae have long been thought to be significant contributors 
to the Galactic abundance of specific isotopes, including by-products of the CNO cycle 
($^{13}$C, $^{15}$N, and $^{17}$O) and those that radioactively decay  ($^{7}$Be, $^{22}$Na, and $^{26}$Al; \citealt{jose98}). 
To quantify these contributions we need to not only understand how these isotopes are created
and how much of them are ejected in individual eruptions,
but also to have a solid estimate of the Galactic nova rate. Take, for example,  $^7$Be---the creation of which has been suggested in nova explosions for decades \citep{arnould1975, starrfield1978}. However, predictions for the amount of $^7$Be created in a typical nova were uncertain \citep{jose98}.
 $^7$Be was recently detected in the ejecta of V339 Del and V5668 Sgr, placing yields on more solid ground \citep{Tajitsu2015, Izzo2015}. 
$^7$Be decays to $^7$Li with a half-life of 53.22 days, so nova eruptions could be responsible for a significant amount of the present day 
Galactic abundance of lithium \citep{romano01,prantzos12,rukeya17,starrfield2020}. 

Another radioactive isotope created during a nova eruption is $^{26}$Al \citep{jose98}. 
This isotope, observed via its
MeV $\gamma$-ray line emission, is also produced in supernovae, 
and has been used as a tracer of the Galactic supernova rate and star formation rate \citep{Diehl06}. 
However, the recent study of \cite{vasini22} shows that novae are likely to be significant contributors to the Galactic $^{26}$Al budget, 
perhaps accounting for the majority of the $^{26}$Al mass (see also \citealt{bennett13}). 
The uncertainty in the nova rate makes determining such nucleosynthetic contributions difficult.

In addition to being an important input to chemical evolution models, 
the Galactic nova rate is also a constraint on binary population synthesis models \citep{Chen16,Kemp21}. 
It has recently been realized that nova eruptions may be an important
mechanism of angular momentum loss from interacting close binary systems
\citep{1998MNRAS.297..633S,2016MNRAS.455L..16S,2022MNRAS.510.6110P}
affecting their evolutionary outcome \citep{2016ApJ...817...69N,2021ApJ...914....5S, Metzger+21}.
In addition, some novae could be the progenitors to Type Ia supernovae \citep[e.g.,][]{patat11, dilday12, darnley15}, depending on the degree to which white dwarfs can retain accreted mass over the course of 
a nova eruption \citep{Toonen14, starrfield2020}. 
Binary models can now reproduce the zoo of accreting white dwarf binaries \citep{Kalomeni16}, and, in the future, a comparison of nova rates with other white dwarf binary populations 
can shed light on how and when white dwarfs manage to grow in mass.

Determining the nova rate of the Milky Way is also important for understanding which---and how many---novae are currently missing from our samples of discovered Galactic novae. Galactic novae are the eruptions we can study in great detail, bringing to bear observations from radio to $\gamma$-ray wavelengths and revealing the physics that drives these eruptions \citep{Chomiuk21}.
However, it is unclear whether the targets of these detailed studies are a representative sample, or if 
particular kinds of novae are missing from our current Galactic samples.

For these reasons, constraining the Galactic nova rate is important, but it has proven to be a tricky feat. Below we summarize the efforts of the  community in discovering Galactic classical novae and the published predictions of the frequency with which the Milky Way produces this class of stellar eruption.

\subsection{History of Nova Discoveries}
Though discoveries of Galactic novae date back millennia, the systematic monitoring of the sky by photographic means for novae began around the turn of the 20th century \citep{Duerbeck08}. During the first half of the 20th century, $\sim$2 novae were discovered per year visually and using photographic plates. With the wide-spread use of astronomical photography \citep{1893AN....134..101P,1895Obs....18..436P} and the advent of objective prism surveys \citep{Duerbeck08}, the rate of discoveries increased to $\sim$3 per year in the mid-20th century. In the 1980s and 1990s, film photography became commonly used by amateur astronomers, and the discovery rate increased to $\sim$4 per year. 

\begin{figure*}
\begin{center}
 \includegraphics[width=1.0\textwidth]{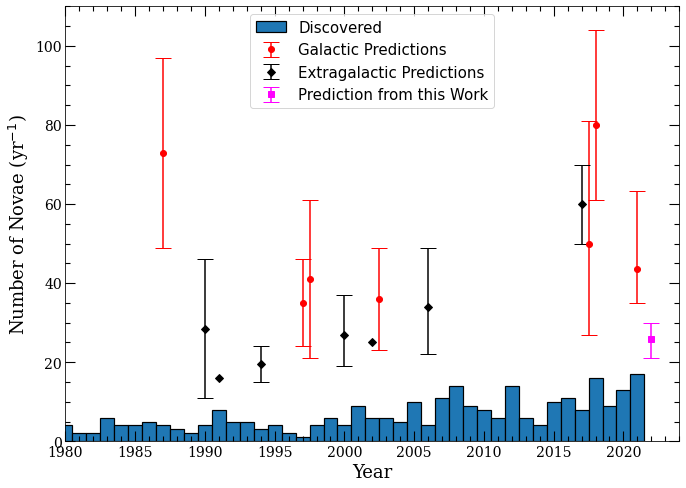}
\caption{Galactic nova rate predictions as a function of the year published, compared with the number of novae confirmed each year since 1980. The rate predictions made using the Galactic or direct method are shown as red circles and using the extragalactic or indirect method are shown as black diamonds, along with error bars if the uncertainty was estimated. The prediction from this work is shown as the magenta square. The number of discovered and confirmed Galactic novae each year is plotted as a blue histogram, and was derived using Koji Mukai's List of Galactic Novae, Bill Gray's Database of Galactic Novae, and AAVSO's VSX. Citations for published rates from left to right: \cite{liller87,cfw90,van91,della94,shafter97,Hatano97,shafter2000,shafter2002,darnley06,shafter_2017,ozdonmez18,de21}.}
\label{fig:rate_hist}
\end{center}
\end{figure*}

In the 2000s and 2010s, more sensitive large-format CCD and CMOS-based digital cameras became widely available to amateur astronomers, 
increasing the annual nova discovery rate to $\sim$8 yr$^{-1}$ (see
Figure~\ref{fig:rate_hist}).
It was thought that discoveries were largely incomplete \citep{liller87}, 
but the degree to which novae were missed due to a shallow magnitude limit, low cadence of observations,
or lack of sky coverage was unknown. Therefore, it was unclear if professional surveys with systematic observations would have a large increase on the discovery rate.

The All Sky Automated Survey \citep[ASAS-3][]{2001ASPC..246...53P} was one of the first CCD surveys imaging the entire sky reachable from its observing site and
contributed to nova discoveries in the early 2000s. Many surveys at the time were focused on 
searching for supernovae, asteroids, 
or transiting exoplanets, so observations of transients in the Galactic plane were lacking, but surveys like ASAS-3 inspired the next generation of wide-field nova searches.

In the 2010s, large sky surveys began to contribute significantly to the discovery of Galactic novae.
The Fourth Phase of the Optical Gravitational Lensing Experiment (OGLE-IV; \citealt{udalski15}) began in 2010, 
with high cadence $I$-band observations of the central bulge
discovering a large number of candidate bulge novae \citep{mroz15}. 
The New Milky Way Survey (NMW; \citealt{sokolovsky14}) began searching for transients in the Northern Hemisphere Galactic plane 
at high cadence down to $V \approx 13.5$ mag in 2011. The All-Sky Automated Survey for SuperNovae (ASAS-SN; \citealt{spg14}) started searching 
for transients in 2013, and then in 2017 became the first survey to systematically observe the entire night sky, including the Galactic plane, 
with nearly daily cadence down to $g \approx$ 18.5\,mag \citep{kss17}. 
These high-cadence and systematic observations allowed for fast-declining novae to be discovered anywhere on the sky. 
The Palomar Gattini-IR survey (PGIR; \citealt{de20}) began surveying the northern hemisphere Galactic plane in 2019 down to $J \approx 15.3$ AB mag. As a near-IR survey, PGIR can discover highly extinguished novae not discovered in previous years \citep{de21}. 
These surveys were designed to discover transients, but the astrometrically focused \textit{Gaia} \citep{gaia16} has been a surprising contributor to the discovery of Galactic novae. 
The broad observing filter, high angular resolution, and all-sky coverage 
allow \emph{Gaia} to detect highly reddened novae anywhere on the sky, including the southern Galactic plane \citep{hodgkin21}. These surveys have helped to increase the average discovery rate to 10.5 yr$^{-1}$ since 2010, with 17 spectroscopically confirmed novae in 2021, the highest number on record (see Figure~\ref{fig:rate_hist}). A higher discovery rate allows for better estimates of the global Galactic nova rate, as there is less sensitivity to model assumptions.

\subsection{Galactic nova rate predictions}
Two methods have been used to estimate the total rate of classical nova production in the Galaxy. The first method extrapolates a sample of discovered Galactic novae based on the estimated completeness (commonly referred to as the Galactic or  direct method). The second method estimates the rate in a nearby galaxy and infers the Milky Way rate by scaling on the relative luminosity of the two galaxies (referred to as the extragalactic or indirect method). 

The direct method was used to make the first prediction for the total frequency of Galactic nova eruptions. It was estimated there should be at least, but probably much higher than, $R =  50$ novae per year \citep{lund35L}. Two decades later, \cite{allen54} used a sample of 19 novae to estimate a Galactic rate of R $\sim 100$ per year.  \cite{Kopylov1955IzKry} used 23 novae thought to be within 1500~pc of the Sun observed
over a 60 year period, arriving at a rate of $R = 50$ Galactic novae per year.
In another two decades,
\cite{sharov72} extrapolated from a sample of eight novae in the Solar neighborhood to estimate a nova rate of $R = 259$ per year for the entire Galaxy. In a similar fashion, \cite{liller87} extrapolated from a sample of 17 novae thought to be within a 60$^\circ$ slice of the Galaxy to a total rate of $R = 73 \pm 24 ~ \rm{yr}^{-1}$. Then, \cite{Hatano97} assumed that novae in the Galaxy are disk dominated and arrived at an annual rate of $R = 41 \pm 21$ yr$^{-1}$.

In the 1990s, estimates began to be derived from extragalactic samples of novae using the indirect method. \cite{cfw90} estimated the nova rate of NGC 5128 to infer the nova rate in our own Galaxy to be between $R = 11-46$ per year. \cite{vandenberghm33} considered the nova rates in M31 and M33 and the globular cluster population ratios to deduce a Galactic nova rate of $R \sim 16 ~ \rm{yr}^{-1}$. \cite{dl94} derived a Galactic rate of $R = 24$ per year by measuring the rates in 5 other galaxies and assuming the rates were proportional to the galaxy luminosity. Finally, \cite{darnley06} used a survey of M31 to infer a Galactic nova rate of $R = 34_{-12}^{+15} ~ \rm{yr}^{-1}$.

\cite{shafter97} is the first in a series of papers that extensively looked at the Galactic nova rate, and by extrapolating samples of novae of varying sector size, limiting magnitude, and time period estimated that the Galactic nova rate is $35 \pm 11~ \rm{yr}^{-1}$. Then, \cite{shafter2000} estimated the nova rates in M51, M87, and M101 to indirectly estimate a Galactic nova rate of $R = 27_{-8}^{+10} ~ \rm{yr}^{-1}$. A couple years later, \cite{shafter2002} assumed that the discovered sample of $m_V < 2$ mag novae are complete, extrapolated to a global rate of $R = 36 \pm 13~\rm{yr}^{-1}$ and also derived a rate of $R \sim25$ per year by comparing the $K$-band luminosity and nova rate of M31 to the Milky Way. In the most recent paper of this series, \cite{shafter_2017} assumed that the $m_V < 2$ mag novae are $90\%$ complete and estimated that the most likely nova rate is $R = 50_{-23}^{+31} ~ \rm{yr}^{-1}$. An even higher rate was predicted a year later when 
\cite{ozdonmez18} estimated the local
density of nova eruptions to predict an average estimate of the disk nova rate of $R = 67_{-17}^{+21} ~ \rm{yr}^{-1}$, and by combining this with a bulge rate estimate of $R = 13.8 \pm 2.6 ~ \rm{yr}^{-1}$ \citep{mroz15}, predicted a Galactic nova rate of $R \sim80 ~ \rm{yr}^{-1}$. 

These predictions from the past 40 years, along with the history of the nova discovery rate, are summarized
in Figure~\ref{fig:rate_hist}.
One noticeable feature is that rate estimates made in the 1990s and early 2000s, dominated by extragalactic predictions, estimated much lower rates than the recent predictions from Galactic data. The \textit{Hubble Space Telescope} (\emph{HST}) survey of M87 novae \citep{shara16} derived a nova rate for M87 over three times larger than that estimated by \cite{shafter2000}. The authors argue that \emph{HST} is more sensitive to faint and fast novae and that previous extragalactic nova surveys had underestimated nova rates because they neglected this harder-to-discover class of novae. Assuming this to be the case, the Galactic rate derived from the M31 sample in \cite{darnley06} was increased to between $\sim 50$ and $\sim70$ per year \citep{shafter_2017}.

Accounting for faint and fast novae could finally make the predictions from both the direct and indirect methods consistent, but it
also appears at odds with the only modest increase in the discovery rate after the advent of large sky surveys: the implication would be that we are still discovering fewer than a quarter of the Galaxy's novae. With these new surveys, it is less plausible that time sampling is the main reason that novae are being missed, so another explanation is needed. A reasonable possibility is dust---that many novae are being hidden from optical surveys by foreground extinction. \cite{kawash21b} quantified the contribution of interstellar dust extinction to the optical discovery rate of Galactic novae and found that dust can hide $\sim50\%$ of the Galactic population from being discovered by observers using $V$- or $g$-band filters. This helps explain some---but not all---of the discrepancy between the discovered and predicted rates.

Luckily, the well defined observing
patterns of large time-domain surveys now make it possible to make systematic predictions of the Galactic nova rate by calculating the expected completeness in the survey data. This new era of Galactic nova rate estimates from large sky surveys began with a prediction using data from the Palomar Gattini-IR survey, where a sample of 11 highly reddened novae was used to derive a rate of $R = 44_{-9}^{+20} ~ \rm{yr}^{-1}$ \citep{de21}. This is consistent with the higher rates that have been published recently, and more predictions from other large sky surveys could help bolster these higher frequency estimates.

For the first time, in this paper we use data from multiple all-sky surveys to estimate the nova rate of the Milky Way. The use of all-sky surveys reduces the sensitivity of our results to possible differences in nova behavior between  the disk and bulge (see, e.g., \citealt{di20})

To date, there are only two all-sky surveys that have reported a nova candidate. First, ASAS-SN became able to scan the entire night sky at a one day cadence in 2017. This high cadence was unprecedented, allowing for the discovery of fast novae anywhere on the sky. 

\textit{Gaia} is the only other all-sky survey 
that reports nova candidates. 
\textit{Gaia} is designed to repeatedly scan the whole sky to make astrometric measurements, and its observing pattern allows for the discovery of many transients, including novae. Usually, a pair of observations are taken 106.5 minutes apart and followed up 2--4 weeks later if the field is not Sun constrained. Though the cadence is much lower than ASAS-SN observations, the high angular resolution ($0.06^{\prime\prime} \times 0.18^{\prime\prime}$), the broad $G$-band observing filter, and the limiting magnitude ($G < 19$ mag.), make \textit{Gaia} sensitive to certain novae in the plane that no other survey can detect \citep{hodgkin21}. So, these two surveys are sensitive to different types of hard to detect novae: ASAS-SN to fast novae and \textit{Gaia} to highly reddened novae in crowded fields. Making a rate estimation using both surveys allows us to better predict the Galactic nova rate, capitalizing on both surveys' distinct strengths.

In Section \ref{sec:disc}, we discuss the sample of novae detected by the surveys and the assumptions we make to calculate the discovery rates. In Section \ref{sec:method}, we explain how we modeled the population of Galactic novae using a Monte Carlo simulation to estimate the Galactic nova rate. Then, in Section \ref{rates} we show the results and compare the simulated detections to the real sample. In Section \ref{sec:compare}, we compare our results to previous estimates and explore how different model assumptions can change the results. Finally, in Section \ref{sec:conc}, we summarize our results and the broader implications.

\section{Discovered Samples} \label{sec:disc}
The direct method of predicting the Galactic nova rate corrects the observed rate for the incompleteness, typically quantified as the recovery efficiency. All spectroscopically confirmed Galactic novae discovered from 2019 to 2021 are listed in Table \ref{table:nova1}. The names and positions of the novae are taken from
an online Galactic nova catalog maintained by one of us (K.M.)\footnote{``Koji's List of Recent Galactic Novae"; \url{https://asd.gsfc.nasa.gov/Koji.Mukai/novae/novae.html}}. Although both ASAS-SN and \textit{Gaia} had been discovering nova candidates before 2019, the observing strategies and pipelines were less stable and hence more difficult to model.

If the nova was detected by ASAS-SN or \emph{Gaia}, we list the date of first detection and the peak detected brightness for each respective survey. The peak brightness does not always occur on the date of first detection, and the peak brightness detected by a survey can be significantly fainter than the true peak brightness if a nova was first detected after a seasonal gap. To be as complete as possible we also list reported transients that were never followed up spectroscopically and could have been classical novae in Table \ref{table:nova2} (individual objects are discussed in Sections \ref{gaia_novae} and \ref{asas_novae}).

\begin{deluxetable*}{lcc|cc|cc|c}
\tabletypesize{\small}
\tablecolumns{8}
\tablewidth{0pt}
\tablecaption{Confirmed Galactic Classical Novae 2019--2021 \label{table:nova1}}
\tablehead{
\colhead{Name} & 
\colhead{RAJ2000} & 
\colhead{DEJ2000} & 
\colhead{ASAS-SN t$_0$} &
\colhead{ASAS-SN peak} & 
\colhead{\emph{Gaia} t$_0$} & 
\colhead{\emph{Gaia} peak} &
\colhead{Ref.}\\
\colhead{ } & 
\colhead{h~m~s} & 
\colhead{$^\circ$~$'$~$"$} & 
\colhead{yyyy-mm-dd} &
\colhead{\textit{g} mag} & 
\colhead{yyyy-mm-dd} & 
\colhead{$G$ mag} &
\colhead{ }}
\startdata
AT 2021abud & 15:27:31.61 & $-$55:06:23.8 & & & 2021-10-06 & 13.9 & \cite{2021abud}\\ 
AT 2021abxa & 17:54:14.14   &  $-$24:12:23.5 & & & 2021-09-20 & 13.8 & \cite{2021abxa}\\
AT 2021aaav & 17:32:21.96 & $-$33:01:41.5 & & & 2021-09-19 & 18.2 & \cite{2021aaav}\\
AT 2021aadi & 18:00:44.88 & $-$21:39:40.5 & & & 2021-09-20 & 16.2 & \cite{2021aadi}\\
ASASSN-21pa & 17:26:19.38 & $-$33:27:10.7 & 2021-07-30 & 17.5 & 2021-08-22 & 14.9 & \cite{21pa}\\
AT 2021wkq & 16:44:50.21 & $-$45:15:48.1 & & & 2021-08-18 & 16.3 & \cite{2021ATel_6novae}\\
RS Oph & 17:50:13.17 & $-$06:42:28.6 & 2021-08-10 & 5.8 & & & \cite{Geary21}\\
V0606 Vul & 20:21:07.70 & $+$29:14:09.1 & 2021-07-16 & 10.6 & & & \cite{V0606Vul}\\
V1711 Sco & 17:39:44.74 & $-$36:16:40.6 & 2021-06-22 & 12.8 & 2021-09-18 & 15.5 & \cite{V1711Sco}\\
V1674 Her & 18:57:30.98 & $+$16:53:39.6 & 2021-06-12 & 9.4 & & & \cite{V1674Her}\\
AT 2021nwn & 19:12:38.61 & $+$12:41:34.4 & & & 2021-05-12 & 15.7 & \cite{AT2021nwn}\\
V2030 Aql & 19:07:58.62 & $+$08:43:45.8 & & & 2021-05-13 & 17.3 & \cite{V2030Aql}\\
V1710 Sco & 17:09:08.11 & $-$37:30:40.9 & 2021-04-12 & 9.7 & & & \cite{V1710Sco}\\
V6595 Sgr & 17:58:16.09 & $-$29:14:56.6 & 2021-04-05 & 9.0 & & & \cite{V6595Sgr}\\
V6594 Sgr & 18:49:05.07 & $-$19:02:04.2 & 2021-03-25 & 10.1 & & & \cite{V6594Sgr}\\
V1405 Cas & 23:24:47.73 & $+$61:11:14.8 & 2021-12-16 & 10.2 & 2021-03-30 & 6.5 & \cite{V1405Cas}\\
V3732 Oph & 17:33:14.83 & $-$27:43:11.0 & & & 2021-02-15 & 15.7 & \cite{V3732Oph}\\
V1112 Per & 04:29:18.85 & $+$43:54:23.0 & 2020-11-26  & 8.9 & 2021-02-16 & 14.1 & \cite{V1112Per}\\
V6593 Sgr & 17:55:00.00 & $-$21:22:40.1 & 2020-09-29 & 11.3 & & & \cite{V6593Sgr}\\
V1708 Sco & 17:23:41.94 & $-$31:03:07.6 & 2020-09-08 & 14.4 & & & \cite{Kojima20}\\
V1391 Cas & 00:11:42.96 & $+$66:11:20.8 & 2020-07-28 & 11.7 & 2020-08-31 & 11.6 & \cite{V1391Cas}\\
V6568 Sgr & 17:58:08.48 & $-$30:05:35.9 & 2020-07-15& 10.6 & 2020-08-14 & 16.9 & \cite{V6568Sgr}\\
YZ Ret & 03:58:29.55 & $-$54:46:41.2 & 2020-07-08 & 6.5 & 2020-08-19 & 7.2 & \cite{YZRet}\\
V2029 Aql & 19:14:26.30 & $+$14:44:40.2 &  & & 2020-08-25 & 14.0 & \cite{V2029Aql}\\
AT 2020oju & 15:25:50.94 & $-$55:10:29.7 & & & 2020-07-09  & 15.9 & \cite{2021ATel_6novae}\\
V6567 Sgr & 18:22:45.32 & $-$19:36:02.2 & 2020-06-02 & 13.5 & 2020-08-18 & 12.8 & \cite{V6567Sgr}\\
V2000 Aql & 18:43:53.33 & $+$00:03:49.4 & & & & & \cite{V2000Aql}\\
V1709 Sco & 17:12:00.18 & $-$40:17:56.7 & 2020-05-10 & 16.1 & & & \cite{V1709Sco}\\
V0670 Ser & 18:10:42.28 & $-$15:34:18.5 & 2020-02-23 & 13.9 & 2020-02-23  & 11.5 & \cite{V0670Ser}\\
V6566 Sgr & 17:56:14.04 & $-$29:42:58.2 & 2020-02-15 & 12.3 & 2020-02-20 & 11.2 & \cite{V6566Sgr}\\
V0659 Sct & 18:39:59.70 & $-$10:25:41.9 & 2019-10-30 & 9.7 & & & \cite{V0659Sct}\\
V2891 Cyg & 21:09:25.53 & $+$48:10:52.2 & 2019-10-21 & 15.6 & 2019-10-07 & 12.7 & \cite{V2891Cyg}\\
V1707 Sco & 17:37:09.54 & $-$35:10:23.2 & 2019-09-14 & 13.2 & 2019-09-15 & 10.1 & \cite{V1707Sco}\\
V3730 Oph & 17:38:31.82 & $-$29:03:47.1 & 2019-09-12 & 16.6$^X$ & 2019-09-14 & 12.3 & \cite{V3730Oph}\\
V3890 Sgr  & 18:30:43.29 & $-$24:01:08.9 & 2019-08-27 & 8.5 & 2019-09-11 & 10.5 & \cite{V3890Sgr}\\
V0569 Vul & 19:52:08.25 & $+$27:42:20.9 & & & 2019-08-24 & 13.5 & \cite{V0569Vul}\\
V2860 Ori & 06:09:57.45 & $+$12:12:25.2 & 2019-08-18 & 12.9 & 2019-08-19 & 11.6 & \cite{V2860Ori}\\
V3731 Oph & 17:38:34.83 & $-$25:19:04.8 &  2019-07-12 & 13.4 & & & \cite{V3731Oph}\\
V1706 Sco & 17:07:34.17 & $-$36:08:23.2 & 2019-05-13 & 13.1 & & & \cite{V1706Sco}\\
N Aql 2019 & 19:03:14.95 & $+$01:20:28.2 & & & 2019-04-06 & 16.2 & \cite{NAql2019}
\enddata
\tablecomments{List of all spectroscopically confirmed Galactic novae discovered between 2019 and 2021. The names and positions are taken from Koji's List of Recent Galactic Novae and AAVSO's VSX. For both ASAS-SN and \emph{Gaia}, the date of first detection $t_0$ and the peak brightness are listed. If no values are listed, it was not detected by that survey. X: Though V3730 Oph was detected by ASAS-SN, it was not flagged and reported in the transient pipeline (likely because it was faint), so it is not included in the ASAS-SN discovery rate. }
\end{deluxetable*}

\begin{deluxetable*}{lcc|cc|cc|c}
\tabletypesize{\small}
\tablecolumns{8}
\tablewidth{0pt}
\tablecaption{Unconfirmed Classical Nova Candidates 2019--2021 \label{table:nova2}}
\tablehead{
\colhead{Name} & 
\colhead{RAJ2000} & 
\colhead{DEJ2000} & 
\colhead{ASAS-SN t$_0$} &
\colhead{ASAS-SN peak} & 
\colhead{\emph{Gaia} t$_0$} & 
\colhead{\emph{Gaia} peak} &
\colhead{Ref.}\\
\colhead{ } & 
\colhead{h~m~s} & 
\colhead{$^\circ$~$'$~$"$} & 
\colhead{yyyy-mm-dd} &
\colhead{\textit{g} mag} & 
\colhead{yyyy-mm-dd} & 
\colhead{$G$ mag} &
\colhead{ }}
\startdata
Gaia21axf & 17:34:38.07 & $-$31:08:00.1 & & & 2021-02-16 & 18.2 & \cite{2021ATel_6novae}\\
Gaia20dfc & 15:22:33.52 & $-$55:59:40.4 & & & 2020-07-09 & 14.9 & \cite{2021ATel_6novae}\\
Gaia20btn & 17:50:19.43 & $-$31:07:37.9 & & & 2020-04-11 & 18.7 & \cite{2021ATel_6novae}\\
ASASSN-19pw & 18:31:05.75 & $-$14:47:52.6 & 2019-06-22 & 15.5 & & & \cite{kawash21}\\
ASASSN-19nf & 14:19:35.09 &  $-$59:58:24.0 & 2019-05-13 & 16.1 & & & \cite{kawash21} \\
ASASSN-19fd & 17:03:19.29 & $-$29:52:23.3 & 2019-03-05 & 13.6 & & & \cite{kawash21}\\
ASASSN-19am & 09:30:39.31 & $-$54:47:04.3 & 2019-01-08 & 16.3  & & & \cite{kawash21}\\
\enddata
\tablecomments{The same columns as Table \ref{table:nova1} for reported transients from between 2019 and 2021 that could have been novae but have no spectroscopic detections. All of these transients are near the Galactic plane, so they could be luminous enough to be novae at a reasonable distance with moderate extinction. The \emph{Gaia} candidates all appear reddened in their $BP/RP$ spectra, but no color information is available for the ASAS-SN transients. These candidates are given $50\%$ weight compared to the spectroscopically confirmed novae when calculating the discovery rate.}
\end{deluxetable*}

\subsection{Gaia Discovery Rate}
\label{gaia_novae}
\textit{Gaia} transients are reported publicly to the \textit{Gaia} science alerts (GSA) website\footnote{\href{http://gsaweb.ast.cam.ac.uk/alerts}{http://gsaweb.ast.cam.ac.uk/alerts}}, but only 25$\%$ of GSA transients have been classified spectroscopically. The vast majority of confirmed candidates are extragalactic supernovae \citep{hodgkin21}, and as seen in Figure 14 of \cite{hodgkin21}, there are a large number of Galactic plane transients that are not classified. Therefore, the number of confirmed classical novae reported by GSA is certainly lower than the actual number detected. To quantify this, we have searched the entire archive of GSA transients for missed Galactic nova events. 

In September 2021, we performed a retroactive nova search on the $\sim17,500$ transients in the GSA website. Those at high Galactic latitude (b$> 4^\circ$) or those with small measured outburst amplitudes (amp. $< 5$ mag) are unlikely to be missed Galactic novae \citep{kawash21}. If the quiescent or historic magnitude is too faint to be detected by \textit{Gaia}, we do not make a cut on the amplitude. We eliminated those at high latitude and those with small amplitudes, and the number of candidates decreased to 435. The bright candidates ($G <$ 14 mag) have all been classified, with the vast majority being classical novae plus a few high amplitude dwarf nova outbursts. The faint candidates ($G >$ 14 mag) also include confirmed, and highly reddened, classical novae, but there are also many unclassified transients. 
In the absence of extinction, an
intrinsically faint nova with peak
absolute magnitude $M_G = -5$ mag at a large Galactic distance of 20 kpc would peak at an
apparent magnitude of $m_G = 11.5$, so the faint ($G >$ 14 mag) nova candidates will all be heavily extinguished and appear red.  Most reported transients also include a low resolution (R $\sim$ 100) and uncalibrated (in wavelength and flux) $BP$/$RP$ spectrum, a tool that has proven to be crucial in identifying highly reddened nova candidates. We examined the spectra of the remaining unclassified candidates to identify any transient with a majority of its flux in the red $RP$ filter or with strong emission lines. This identified five highly reddened Galactic plane transients detected between 2019 and 2021 with no spectroscopic followup. 

We obtained spectra of these candidates using the Goodman spectrograph \citep{Clemens_etal_2004} on the 4.1\,m Southern Astrophysical Research (SOAR) telescope to look for evidence of past nova eruptions. We detected spectral features in Gaia20dfb and Gaia21dwe consistent with past nova eruptions \citep{2021ATel_6novae} and list them as confirmed novae in Table \ref{table:nova1}. Because the followup observations were taken months to years after the eruption and the sources are likely highly extinguished, we did not detect flux from Gaia21axf, Gaia20btn, and Gaia20dfc, but we still consider these transients as Galactic nova candidates and list them in Table \ref{table:nova2} because their positions, colors, spectral features, and lightcurves are all consistent with other classified classical novae. Their colors rule out the most common type of classical nova contamination (dwarf novae;  \citealt{kawash21}), but the chance of contamination from microlensing events, young stellar objects, and Be star outbursts remains.
Given the rate of confirmation of other candidates, and in lieu of a more complex contamination model, we assign each of these 
candidates a 50\% probability of being a nova for our Monte Carlo simulations described in Section \ref{sec:method}.

As seen in Table \ref{table:nova1}, \textit{Gaia} reported 7, 8, and 12 confirmed novae in 2019, 2020, and 2021, respectively, with an additional 3 unconfirmed candidates listed in Table \ref{table:nova2}. Assuming Poisson uncertainty for this sample of $\sim 28.5$ novae yields a mean and standard deviation of the discovery rate of 9.5 and 1.8 yr$^{-1}$, respectively when running the Monte Carlo simulation over the years 2019 to 2021. The dates listed as ``\emph{Gaia} t$_0$" in Table \ref{table:nova1} are the epoch of \emph{Gaia}'s first detection, which can lag the start of the eruption by weeks to months because of \emph{Gaia}'s non-uniform scanning law.

\subsection{ASAS-SN Discovery Rate}
\label{asas_novae}
The ambitious goal of ASAS-SN to observe the entire night sky daily has provided unprecedented cadence for Galactic observations of transients. ASAS-SN transients are reported publicly to \href{https://www.astronomy.ohio-state.edu/asassn/transients.html}{https://www.astronomy.ohio-state.edu/asassn/transients.html}. In 2018, ASAS-SN switched from observing in $V$-band to exclusively observing in $g$-band. To allow time for deep $g$ band reference images to be built, we only calculate the discovery rate between 2019$-$2021. Over this time period we inspected all the ASAS-SN data of confirmed Galactic novae, and found that 26 novae were detected in the transient pipeline. If there is no ASAS-SN value for $t_0$ or peak brightness listed in Table \ref{table:nova1}, the nova was too highly reddened, and therefore too faint ($g \lesssim 18.5$ mag) to be detected by ASAS-SN. V3730 Oph was detected at $g = 16.6$ mag but was never flagged in the pipeline as a transient (fainter transients have a smaller chance of being reported relative to brighter ones). We therefore do not include it as a nova in the ASAS-SN discovery rate. \cite{kawash21} found that there were four more reported CV candidates in 2019 that could be luminous enough to be novae if they are distant enough to be significantly obscured by dust extinction. As for the \emph{Gaia} candidates, we make the simple assumption that these four candidates, listed in Table \ref{table:nova2}, have a 50\% probability of being classical novae. So, over the three years considered here, there were 26 confirmed and 4 unconfirmed novae, for a total population of $28$ novae. Assuming Poisson statistics yields a mean and standard deviation of the ASAS-SN discovery rate of 9.3 and 1.8 yr$^{-1}$, respectively. 

\subsection{The joint ASAS-SN and Gaia Discovery Rate}
If we combine both surveys and account for overlapping discoveries, there were a total of 39 confirmed classical novae and 7 additional candidates detected between 2019--2021. This yields a mean and standard deviation of the discovery rate of 14.2 and 2.3 yr$^{-1}$, respectively. While this observed nova rate is still far below most predictions, it is still the highest \emph{observed} rate ever used to infer the total nova rate.

\section{Monte Carlo Simulations}
\label{sec:method}
Our work attempts to repeatedly answer this basic question: if a nova erupted at a specific time and location in the Galaxy, would it be detected and reported as a transient by ASAS-SN and/or \textit{Gaia}? This idea is expanded to a large sample of simulated novae to estimate what fraction of the Galaxy's novae these surveys detect. By implementing a Monte Carlo analysis, we derive the most likely Galactic nova rate and accompanying uncertainty based off of ASAS-SN and \textit{Gaia} observations and the uncertainty in the model parameters. 

\subsection{Positions}
The positions of the model novae in the Galaxy are derived by assuming that they trace the stellar mass density of the Galaxy. The stellar density model is outlined in the appendix of \cite{kawash21b} and includes a thin disk, thick disk, and halo component as described in \cite{robin03} and a two component, elongated, triaxially symmetric bulge from \cite{sbi17}. The ratio of the disk mass to the bulge mass, and therefore the ratio of disk to bulge novae assuming equal production per stellar mass, is 1.7. For each run of the Monte Carlo simulation, we randomly pick 1,000 nova positions from a possible 10,000 positions. Because this study requires lightcurves to be generated in each survey at every position, we limit the total number of possible positions to 10,000. In \cite{kawash21b}, we explored how sensitive the derived nova rate is to the stellar density model, where it was found to change results by $\sim~15\%$ for the ASAS-SN rate, subdominant to other uncertainties in our calculation.

\subsection{Peak Apparent Magnitude}
\label{sec:peak_mag}
The peak absolute magnitude of each nova is assigned by randomly sampling a normal distribution, described with a mean and standard deviation of $M_{g,G} = -7.2 \pm 0.8$ mag \citep{shafter_2017}. This distribution was measured from M31 novae, and we assume the Milky Way has an identical distribution.

The extinction along the line of sight to each nova position is calculated from the \texttt{mwdust} package \citep{bovy15}. We use the \texttt{combined19} version of this extinction model, which is built by combining the \cite{marshall06} map of the inner Galactic plane, the \cite{green19} map of the Northern Hemisphere, and the \cite{drimmel03} map of the Southern Hemisphere. The various maps take precedence over each other where they overlap in the order they were listed above. The vast majority of the model novae ($95\%$) lie in the \cite{marshall06} map region of the sky. For ASAS-SN, we query the extinction in the SDSS $g$-band, as it is the most similar to ASAS-SN's $g$-band filter. Extinction in the broad \textit{Gaia} $G$-band filter is not directly accessible in \texttt{mwdust}, so we estimate this value by first querying the extinction in the SDSS $g$- and SDSS $i$-bands. The relationship between these SDSS filters and \textit{Gaia's} $G$-band filter were fit to a third order polynomial with the form 
\begin{equation}
    G-g = -0.1064 - 0.4964x - 0.09339x^2 + 0.004444x^3
\end{equation}
where $x = g-i$ \citep{gaia_edr3}. 

\begin{figure}
\begin{center}
 \includegraphics[width=0.45\textwidth]{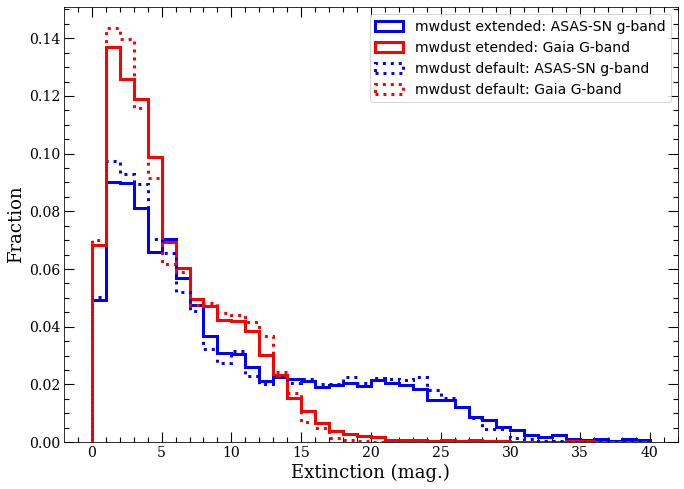}
\caption{The normalized extinction distributions of 10,000 simulated novae in the ASAS-SN $g$-band filter (blue) and \emph{Gaia} $G$-band filter (red). These distributions are estimated from the \texttt{mwdust} package (default values shown as dotted lines), and we add additional extinction if the line of sight is not complete out to the distance of the nova (solid line).}
\label{fig:ext_hist}
\end{center}
\end{figure}

An additional concern when estimating the  extinction for a model of Galactic novae is that the 3D dust maps that comprise the \texttt{mwdust} package only estimate the extinction out to distances where the colors of stars can be measured, and therefore do not extend to the largest Galactic distances and the highest extinction regions. If a model nova is beyond the largest distance bin of \texttt{mwdust}, we add extinction by fitting the amplitude of the extinction along a line of sight and assuming the dust is described with an underlying 
double exponential distribution that is a function of Galactic radius and height from the plane. We use ($r_0,z_0$) $=$ (3.0,0.134) kpc for the scale length and height, respectively \citep{Li_2018}. The amount of extinction is fit out to the distance where \texttt{mwdust} has information, then we use the results to extrapolate to the nova distance. For $55\%$ of simulated novae, \texttt{mwdust} extends to the distance of the nova, so no extinction is added. For those remaining novae beyond the 3D dust maps, the mean extinction added in the $g$- and $G$-bands is $0.9$ and $0.6$ mag, respectively. The distributions of estimated extinction before and after this correction are shown in Figure~\ref{fig:ext_hist}. The median nova will experience $\sim10$ mag of extinction in $g$-band and $\sim6$ mag of extinction in $G$-band.

\begin{figure}
\begin{center}
 \includegraphics[width=0.45\textwidth]{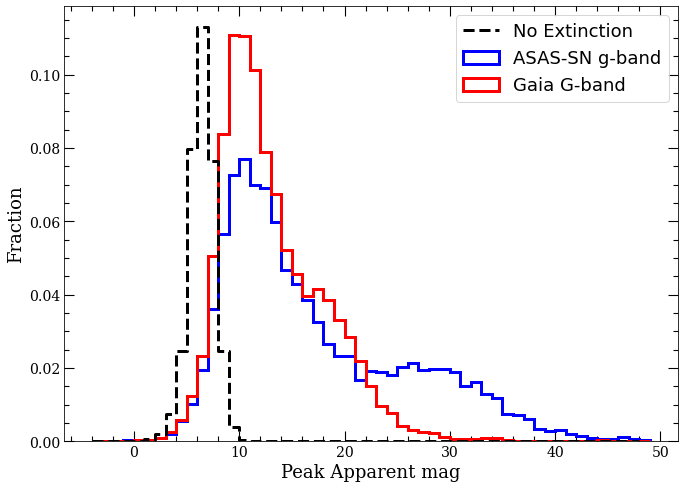}
\caption{The normalized distributions of peak apparent magnitude of 10,000 simulated novae in the $g$-band filter (blue) and $G$-band filter (red). Also shown is the distribution with zero extinction with the peak scaled to fit (black dashed line). This demonstrates that dust is a much larger factor than the luminosity or distance in determining the peak apparent magnitude of a nova.}
\label{fig:peak_hist}
\end{center}
\end{figure}

With the peak luminosity, the distance, and the extinction in $g$- and $G$-bands estimated, the peak apparent magnitude of each model nova can be calculated using the distance modulus. The results are shown for 10,000 model novae in Figure~\ref{fig:peak_hist}. The differences between the \emph{Gaia} and ASAS-SN peak apparent magnitude distributions are entirely due to the different observing filters used; the bluer ASAS-SN $g$-band is more vulnerable to extinction from interstellar dust compared to the broader \textit{Gaia} $G$-band. This figure shows that extinction is the largest factor in determining the apparent brightness of a Galactic nova. The dashed black distribution ignores extinction, and therefore its variance is caused by variations in the luminosity and distance of a nova; in this case, the median absolute deviation in the peak apparent brightness is only 1.1 mag. However, when considering extinction, the median absolute deviation of the peak brightness seen in \emph{Gaia}'s $G$-band filter is 4.2 mag, and in ASAS-SN's $g$-band filter, the median absolute deviation is 7.2 mag. Based only on the peak brightness, $\sim 90\%$ percent of Galactic novae are bright enough to be detected by \textit{Gaia} ($G < 19$ mag), compared to the only $\sim 60\%$ percent detectable by ASAS-SN ($g < 18.5$ mag.).

Some studies have suggested that Milky Way novae are more luminous than M31 novae \citep{shafter09,ozdonmez18}, but we do not explore different luminosity functions as it is clear that extinction is a much larger factor than the absolute magnitude for determining the brightness of a nova in the optical.

\subsection{Lightcurves}
The modeling work to derive the apparent magnitude distribution of Galactic novae was largely laid out in \cite{kawash21b}, but the detection efficiency in the survey data was only broadly estimated. Here, we incorporate the model into each survey's data to more accurately estimate the detection efficiency. Each simulated nova is given a random eruption date between January 1, 2019 and December 31, 2021. While in principle novae first discovered at the start of our survey period (2019 January) could have exploded in 2018, and hence outside of our simulation, the predicted and observed number of novae in the first half of January is so low that this potential issue does not have a meaningful ($\lesssim 1\%$) affect on our results. Though not entirely symmetric, this effect is also diminished by novae that erupt near the end of the simulation in late 2021 that would be discovered in 2022, after the end of the simulation. 

The temporal evolution of the brightness is modeled by sampling a distribution of nova speeds and then constructing the full shape of the lightcurve from a sample of known nova lightcurves. We used 93 AAVSO $V$-band nova lightcurves from \cite{ssh10} and 75 Stony Brook/SMARTS $V$-band lightcurves from \cite{walter12} as lightcurve templates. $V$-band was chosen because it was the most well-sampled filter from the two databases and also close to the $g$ and $G$ bands of ASAS-SN and \emph{Gaia}. Motivated by the typical intrinsic colors of novae (e.g., \citealt{1987A&AS...70..125V}), we assume a flat spectrum ($V_{\rm{peak}}  = g_{\rm{peak}}  = G_{\rm{peak}}$) when transforming these templates to $g$- and $G$-band.

Previous completeness studies have used several methods to estimate how long a nova is detectable, including defining a discrete number of observable days after eruption \citep{mroz15} and assuming a linearly declining lightcurve \citep{de21}. By using real lightcurves as templates, we do not have to make simplifying assumptions about the shape of the lightcurves, which can be quite complex \citep{ssh10}. However, we also carry out the analysis by using linearly declining lightcurves to explore how sensitive the results are to this property and discuss the results in Section \ref{sec:pgir}.

We took additional steps to reduce errors in the light curve templates. A template should only include flux from the nova eruption and not from when the nova has returned to the quiescent state or from nearby background stars. Most of the quiescent magnitudes are listed in Table 1 of \cite{ssh10} for the novae presented in that work, so we truncate the lightcurve once it declines to within 3 mags of this quiescent magnitude. \cite{walter12} notes that there are cases in the SMARTS sample of novae when the fading remnant becomes blended with a background star, and from inspecting these lightcurves, this does appear to be true in a few cases. We also cut off lightcurves within a magnitude of any prolonged plateaus many magnitudes below maximum to eliminate this contamination. If the duration of the template lightcurve is shorter than the survey length, we extend the template by assuming it will decline linearly at the average pace of the lightcurve.

Each nova is randomly assigned a decline speed by sampling a log-normal distribution for t$_2$ (the time it takes a nova to decline two magnitudes from peak) with mean and standard deviation equal to 18.7 and 3.2 days, respectively, based on modeling of observed novae and accounting for selection effects \citep{kawash21}. 

\begin{figure}
 \includegraphics[width=0.5\textwidth]{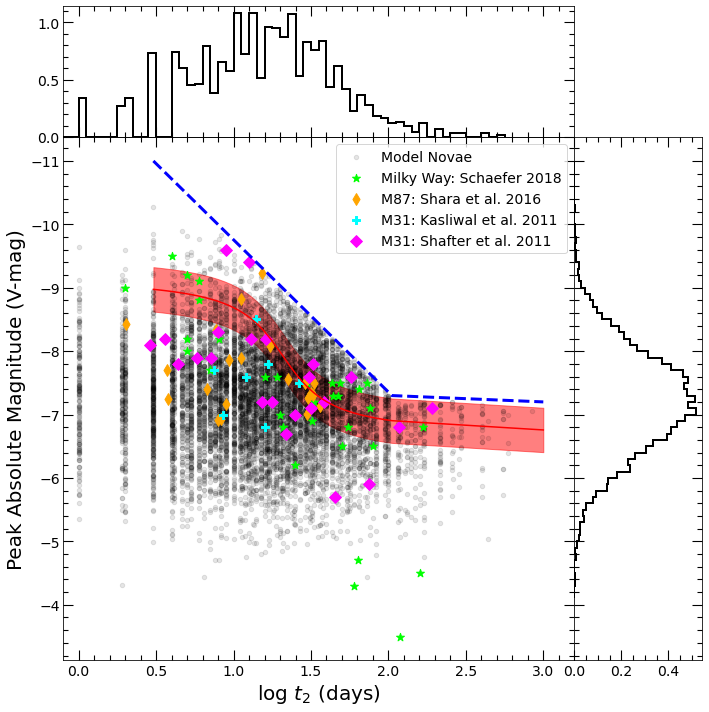}
\caption{Distribution of the peak absolute magnitude versus time to decline by two magnitudes from maximum (t$_2$) for 10,000 model novae (black dots). The peak absolute magnitudes are sampled from a normal distribution, and the t$_2$ values are sampled from a log-normal distribution. The patchiness in the t$_2$ distribution is the result of a limited number of nova lightcurve templates. The allowed values in this parameter space are shown below the blue dashed line, and this is compared to real Galactic nova values estimated from \emph{Gaia} distances (\citealt{Shaefer18}; denoted with green stars), extragalactic measurements (pink diamonds, cyan crosses, and orange diamonds; \citealt{shafter_2011,kck11,shara16}), and the MMRD correlation derived in \cite{DellaValle&Livio95} (shown as the red shaded region).}
\label{fig:mmrd}
\end{figure}

There has been a large effort to establish novae as standard candles. The relationship between the luminosity and the speed of a nova, commonly referred to as the maximum magnitude vs.~the rate of decline (MMRD) relation, was once thought to be tightly correlated \citep{Capaccioli+89, DellaValle&Livio95}, but in recent years, many examples of novae that do not follow the published relationships have been found, especially those that are fainter and faster \citep{kck11,sdl17}. These faint fast novae have been suggested to be a reason for a factor of three discrepancy in extragalactic nova rate estimates \citep{shara16}. In our primary model, we assume the luminosity of a nova simply sets an upper bound on the speed or t$_2$ value. Put another way, the luminous novae are restricted to having small t$_2$ values, but the fainter novae are allowed all values of t$_2$. The boundary between the forbidden and allowed values is shown as a blue dashed line in Figure~\ref{fig:mmrd} with the allowed values being below this line. Any model nova with a given luminosity that is assigned a forbidden t$_2$ value has this parameter resampled until an allowed value is found. This forbids luminous slow nova, as no such example has been found but allows for any number of faint fast novae. 

\begin{figure*}
\begin{center}
 \includegraphics[width=1.0\textwidth]{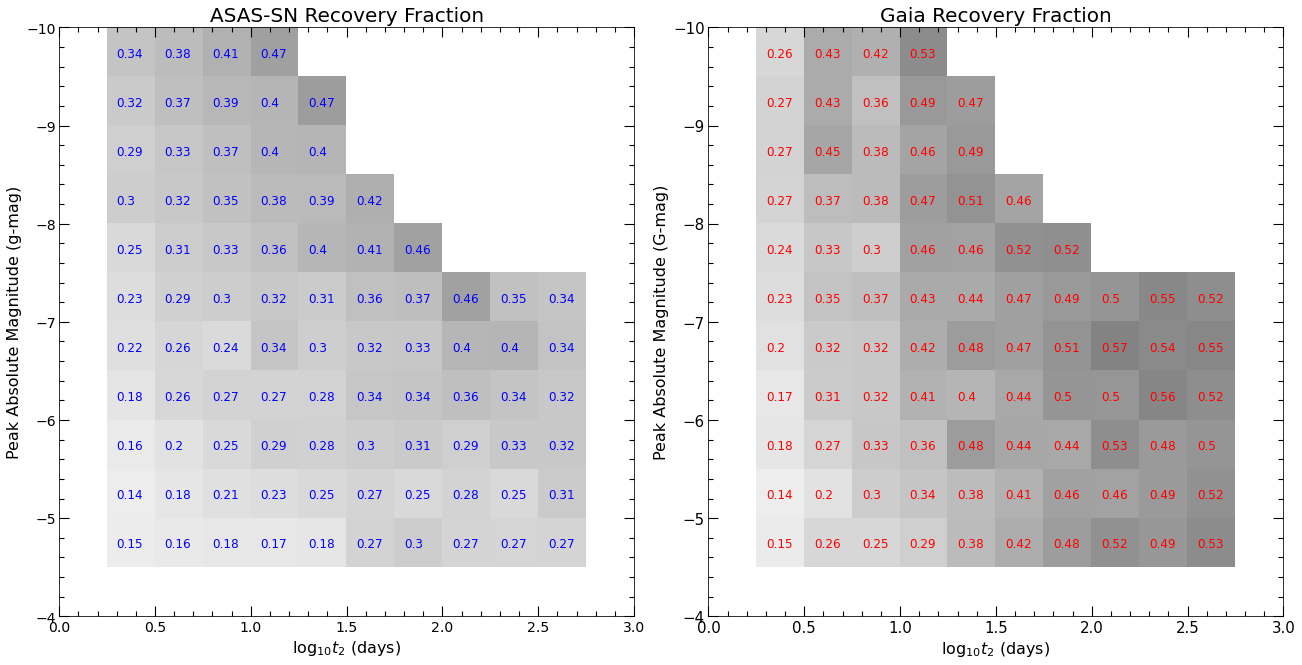}
\caption{The recovery fraction of model novae in ASAS-SN and \emph{Gaia} as a function of peak absolute magnitude and $t_2$. The recovery fraction values are calculated on a grid of 0.5 mag by 0.25 $\log_{10}(t_2)$ days, and are shown on each square, with darker grey-scale indicating a higher recovery fraction. While novae that adhere to the supposed MMRD relation are on average easier to detect, they are only recovered about twice as efficiently as faint fast novae, implying that all-sky surveys would have detected a substantial population of faint fast novae if they were present.}
\label{fig:mmrd_sim}
\end{center}
\end{figure*}

The maximum magnitude versus rate of decline of 10,000 model novae are shown in Figure~\ref{fig:mmrd}. The distribution of t$_2$ is discontinuous because of the discrete t$_2$ values of the template lightcurves. Also show on this plot is the MMRD correlation measured in \cite{DellaValle&Livio95} as the red shaded region,  real Galactic nova values measured from ``Gold" and ``Silver" \textit{Gaia} distances (denoted as the green stars; \citealt{Shaefer18})
, and various extragalactic novae \citep{shafter_2011,kck11,shara16}. The faint, fast novae are arguably over represented in our model compared to observations, so we also rerun the analysis by treating MMRD as a strict correlation and discuss the results in Section \ref{sec:pgir}. 
We explored the detection efficiency of each survey in this parameter space (see \S \ref{sec:gaia} and \ref{sec:asassn} for details on detection efficiency), and the results are shown in Figure \ref{fig:mmrd_sim}. Though faint fast novae are clearly harder to detect in the model, the difference in the recovery fraction of faint fast novae compared to novae that adhere to the MMRD relation is no more than a factor of two. This means that it is unlikely a large population of faint fast novae exists in the Galaxy that has not been detected by all-sky surveys, unless such novae have a very different spatial distribution and are more embedded in dust than typical novae.

Once each model nova is given a decline rate, we assign a lightcurve template by finding the \cite{ssh10} or \cite{walter12} lightcurve closest to the randomly assigned t$_2$ value. The template is then scaled so that the peak apparent magnitude matches the value calculated above. The model novae now have all of the information needed to inject them into the survey data, and below we discuss how that is performed for each survey. 

\subsection{Gaia Simulation}
\label{sec:gaia}
\begin{figure*}
\begin{center}
 \includegraphics[width=1.0\textwidth]{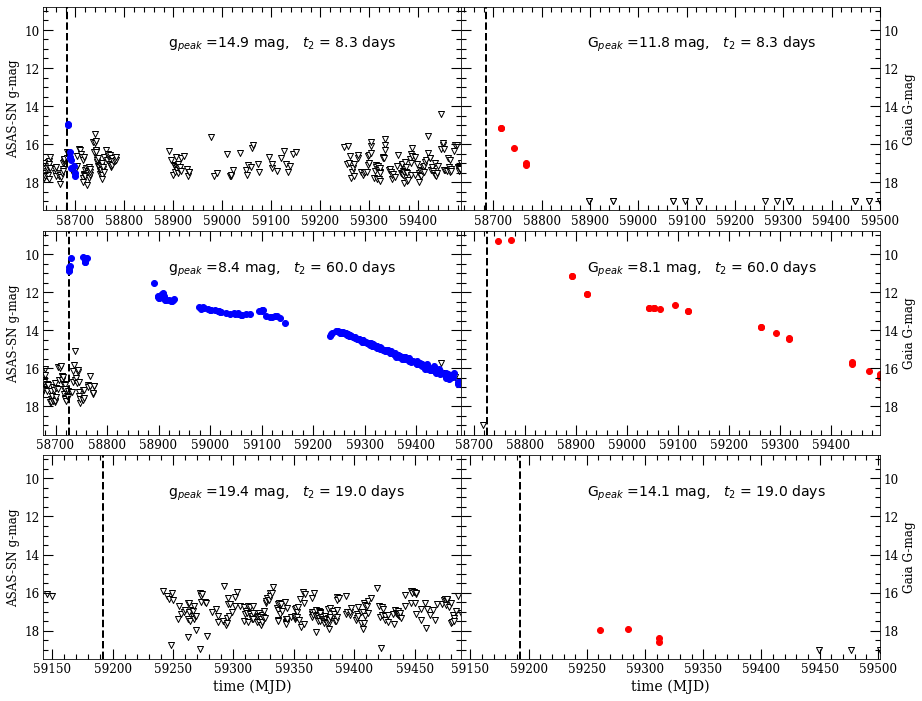}
\caption{Three examples of simulated novae in ASAS-SN (left) and \textit{Gaia} (right). The detections are shown in blue for ASAS-SN and red for \textit{Gaia}, with non-detections shown as black triangles. The epoch of eruption is denoted by the vertical dashed line and the peak apparent magnitude and $t_2$ are listed in each panel. The top row shows a fast reddened nova, the middle row shows a nearby and slow nova (the non-detections shortly after the eruption indicate that the transient has saturated the ASAS-SN detectors), and the bottom row shows a reddened nova that was only detected by \textit{Gaia}.}
\label{fig:nova_in_data}
\end{center}
\end{figure*}

For each nova position we used the Gaia Observation Forecast Tool\footnote{\href{https://gaia.esac.esa.int/gost/}{https://gaia.esac.esa.int/gost/}}
to provide the epochs at which \textit{Gaia} observed the location, and we assume a fixed detection threshold of $G <$ 19 mag. We then sample the template lightcurve at the cadence of \emph{Gaia}, and several examples are shown in the right-hand column of Figure~\ref{fig:nova_in_data}. We draw the noise in the lightcurve from a normal distribution with standard deviation
\begin{multline}
    \sigma =  3.43779 ~ \rm{mag} - (G/1.13759 ~  ) + \\
    (G/3.44123~ )^2 - (G/6.51996~ )^3 +
    (G/11.45922~ )^4
\end{multline}
if $13 < G < 19$ mag and $\sigma = 0.02$ mag for $G < 13$ mag.

To determine if a model nova would be reported by GSA we start with three basic requirements (outlined in \citealt{hodgkin21}). First, there need to be detections in both FOVs (fields of view; preceding or trailing) brighter than $G < 19$ mag within 40 days of each other. Second, the detected brightness needs to exceed that of all stars within a 1.5$^{\prime\prime}$ radius in \textit{Gaia} DR2, and last, there cannot be a $G < 12$ mag star within a 10$^{\prime\prime}$ radius in \textit{Gaia} DR2. 

Even if a transient satisfies these three requirements, there are additional ways for it to be missed in the pipeline. For example, a single source can have: multiple source IDs, ambiguous matches in \emph{Gaia} DR2, or multiple sources within the core region, etc. \citep{kostrzewa18}. It is difficult to determine when this will happen in the simulation, so we make a statistical estimate by looking at the reporting efficiency of novae in Table \ref{table:nova1}. There are 20 novae reported by GSA that were also detected by other surveys (amateurs, ASAS-SN, PGIR, etc.); however, after further inspection of the \textit{Gaia} database and the \textit{Gaia} observation forecast tool, there are 7 additional novae (V0569 Sct, V1708 Sco, V5693 Sgr, V1710 Sco, V1674 Her, V0606 Vul, and RS Oph) that should have been reported based on the three major criteria listed above. So, GSA reported no more than $20/27 \approx 74\%$ of the novae that passed these criteria, and the true reporting efficiency is likely somewhat lower because of the strong chance of additional candidates that were unreported by other observers. We treat this estimate as a $1\sigma$ upper limit with $10\%$ uncertainty, so for each Monte Carlo trial, we assign \emph{Gaia} a reporting probability for novae that pass all of the hard-coded requirements by sampling a normal distribution with mean $\mu = 0.67$ and standard deviation $\sigma = 0.067$. This is necessarily a crude model to summarize the complex process that leads to a candidate nova being reported and undoubtedly an important source of uncertainty in our simulations. Future observations of novae and \emph{Gaia} alerts can help constrain this reporting efficiency. 

\subsection{ASAS-SN Simulation} \label{sec:asassn}
The cadence and limiting magnitude of ASAS-SN observations were calculated by constructing image subtraction lightcurves without adding the reference flux from the ASAS-SN data over the 10,000 positions of the model novae. This automatically provides a sampling of the cadence, noise uncertainty (due to the lunar cycle and weather conditions), and contamination from bright, nearby stars.  
The formal photometric uncertainties reported in ASAS-SN lightcurves tend to underestimate the true uncertainties \citep{jaya18}, and this can particularly be true in the Galactic plane due to crowding, even with the benefits of image subtraction. We estimated a rescaling of the uncertainties for each light curve by looking at the distribution of the ratio $f_i/\sigma_i$ where $f_i$ is the flux and $\sigma_i$ is the reported error in the flux of each camera at each position. If the error estimates are correct, then the standard deviation of this distribution should be unity. When it is larger, contamination is present, so we increase $\sigma_i$ by the factor needed to make the distribution unity. For many of the random nova light curves, the rescaling is large enough to lead to a $\sim 1$~mag reduction in
sensitivity on average. Because this work is looking at Galactic novae which tend to be located at crowded low Galactic latitudes, these affects are more severe than compared to an extragalactic supernova study. 

When injecting the model nova lightcurves into the ASAS-SN data, we assume that the nova would be detected if it is brighter than these rescaled $5\sigma$ upper limits on each particular epoch. Similar to the \textit{Gaia} analysis, we fit a polynomial to photometric errors in ASAS-SN data, and we add noise to the lightcurve templates by sampling a normal distribution with a measured standard deviation of
\begin{multline}
\sigma = 
0.08~ \rm{mag} + 0.04(g-13) - 0.04(g-13)^2/\rm{mag} ~ + \\ 0.02(g-13)^3/\rm{mag}^2 - 0.002(g-13)^4/\rm{mag}^3
\end{multline}
if $13 < g < 18.5$ mag, and $\sigma = 0.02$ mag for any $g < 13$ mag.

When a transient is reported by ASAS-SN, it is cross checked to catalogs like \textit{Gaia} and the Panoramic Survey Telescope and Rapid Response System (Pan-STARRS; \citealt{cmm16}) to roughly estimate the outburst amplitude. This helps differentiate classical nova candidates from other CV outbursts, but the large pixel scale of ASAS-SN (8$^{\prime\prime}$ per pixel) can lead transients to have underestimated outburst amplitudes when their position is coincident with another star. To account for this, we also require detections to be 5 mags brighter \citep{kawash21} than the closest star in \textit{Gaia} DR2 within half an ASAS-SN pixel, to assure the transient would be recognized as having a large outburst amplitude. 

The goal of the ASAS-SN pipeline is to discover previously unknown transients and variable stars. To avoid continually looking at known variables, the ASAS-SN pipeline does not generate candidate images for flux changes detected within 5 pixels of known Mira, long-period, or semi-regular variables. A list of the positions, types, and magnitude range of known variables is acquired from VSX and OGLE and is maintained in the ASAS-SN database.  We inspect this same list in the simulation and require that the model novae also satisfy this condition. If a transient is near a different type of variable, a candidate image will be generated; however, if a new transient is found near a known one, it is possible for the new transient to be confused with a known repeater and consequently not reported. To incorporate this into the model, a transient has to be a magnitude brighter than the listed magnitude range of any known variable within half an ASAS-SN pixel of the simulated nova. 

The last step for a transient that is detected by ASAS-SN to be reported is that a human needs to flag the source in the data as real and previously unreported. As mentioned above, the probability of this occurring scales with the SNR. The images of bright candidates ($g < 15$ mag) generate alert emails and are filtered into a pipeline dedicated to finding Galactic classical novae. Fainter Galactic transients are commonly reported by ASAS-SN but with lower completeness due to the number of false positives from artifacts increasing at fainter magnitudes. Generally, the closer a candidate is to the detection limit, the less likely the image is to be vetted by a human and reported as a transient. The probability for a single detection of an extragalactic supernova to be reported in ASAS-SN was studied
by {\bf Desai et al. (2022, in prep)} and found to be 
\begin{equation}
    \left\{\begin{matrix}
0.65 & SNR >= 12 \\ 
(1 + (\rm{SNR}-12)/7)\times 0.65 & SNR < 12
\end{matrix}\right.
\end{equation}
where SNR is the signal-to-noise ratio of each detection. In the simulation, each detection that satisfies the above requirements has this probability of being flagged. This detection probability is per individual epoch, so the brighter and slower transients have a much higher likelihood of being reported than the faint and fast ones.

\section{Global Nova Rate Estimates}
\label{rates}

\begin{figure*}
\begin{center}
 \includegraphics[width=1.0\textwidth]{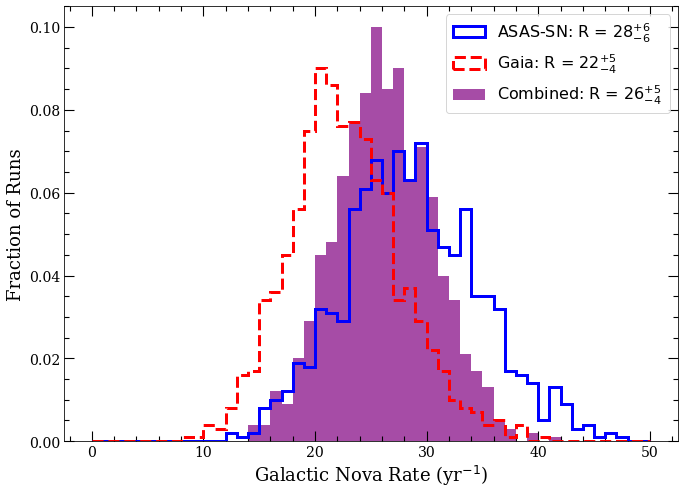}
\caption{The predicted Galactic nova rate based on 1,000 Monte Carlo trials, from ASAS-SN (blue histogram), \emph{Gaia} (red histogram), and a combination of both surveys (purple histogram). We give the median values of the distributions as the most likely Galactic nova rate, and include the $16\%$ and $84\%$ confidence regions. The redundancy of the three distributions is low, as the combined results are sensitive to when a nova is detected by both, just one, or neither surveys. The results are all consistent at the $1\sigma$ level, with the most likely rate from both surveys predicting a Galactic nova rate of $R = 26 \pm 5$ yr$^{-1}$.}
\label{fig:rates}
\end{center}
\end{figure*}

The results of the Monte Carlo simulation show the probability distribution of the global Galactic nova rate based upon ASAS-SN and \textit{Gaia} observations and reporting of Galactic novae between 2019 and 2021.
We run the Monte Carlo 1,000 times, each time sampling different values for: (i) the nova positions, (ii) the lightcurves' speed and shape, (iii) eruption dates, (iv) peak luminosities, (v) the probability of detected transients being reported, and (vi) the discovery rates based on Poisson sampling. The Galactic nova rate is then calculated for each trial by dividing the discovered rate by the estimated recovery fraction for each respective sample. Each trial provides a different estimate of the Galactic nova rate, and we take the median rate as the most likely and $68.2\%$ of the width as the $1\sigma$ uncertainty. The results of the Monte Carlo simulation are shown in Figure~\ref{fig:rates}.

We simulate the \textit{Gaia} and ASAS-SN surveys individually and as a joint search, providing three estimates of the Galactic nova rate. The joint estimate might appear to be redundant with the former two, but the fraction of novae discovered by neither, one, or both surveys provides additional information not captured in the prediction from a single survey since the surveys have different filters and cadences. The three rates are all consistent at the $1\sigma$ level: ASAS-SN, \textit{Gaia}, and the combination of both surveys predict global Galactic nova rates of $28_{-6}^{+6} ~ \rm{yr}^{-1}$, $22_{-4}^{+5} ~ \rm{yr}^{-1}$, and $26_{-4}^{+5} ~ \rm{yr}^{-1}$, respectively. This work estimates that the combined efforts of both all-sky surveys were able to detect  $\sim54\%$ of the Milky Way's classical nova eruptions that occurred between 2019$-$2021, a much higher recovery fraction than recently estimated \citep{ozdonmez18,de21}. Individually we estimate that the recovery fraction of ASAS-SN is $\sim 33\%$ and the recovery fraction of \textit{Gaia} is $\sim 42\%$.

Breaking those estimates down further, for ASAS-SN, about $40\%$ of novae are too faint for discovery because they are too highly extinguished, an additional $\sim 20\%$ are lost because there is not an observation while the nova is brighter than the average detection limit of $g < 17$ mag (largely from seasonal gaps), and the last $\sim 7\%$ are lost because of various pipeline features (low SNR, confusion, avoiding known variables, etc.). This is a lower fraction of novae lost because of cadence and the pipeline than the assumption of $40\%$ used in \cite{kawash21b}, so the nova rate derived in that work was overestimated. 

For \textit{Gaia}, only $\sim 14\%$ of novae are too faint because they are too highly extinguished, $\sim 12\%$ are lost because of lack of cadence, and $\sim 35\%$ are lost because of various pipeline features (requiring detections in both FOVs, source confusion, etc.). This is consistent with the analysis of extragalactic supernovae that found that the scanning law and the need to minimize the false alarm rate dominates the completeness of GSA \citep{hodgkin21}. Surprisingly, the higher cadence of ASAS-SN loses more novae that peak bright enough for detection than \textit{Gaia}, but this is again because of dust extinguishing novae so that the bluer ASAS-SN observations have a much shorter time for discovery and because ASAS-SN has a lower recovery rate at fainter magnitudes. 

To assess the accuracy of the model, we compare the results of the simulation to the real novae in Table \ref{table:nova1}. Any significant conclusions should be taken with caution because of the small sample size of discovered novae, but this exercise can still shed light on the accuracy of the model. First we look at what fraction of the survey's discovery sample were also discovered by the other survey. In the simulation, about $60\%$ of the ASAS-SN discovered novae were also discovered by \textit{Gaia}, compared to the observed value of $14/26 \approx 54\% \pm 14\%$ (assuming Poisson uncertainty) of the ASAS-SN sample from Table \ref{table:nova1} that were detected by \textit{Gaia}. Roughly 50$\%$ of the simulated novae discovered by \textit{Gaia} were also discovered by ASAS-SN, compared to the observed $14/30 \approx 47\% \pm 12\%$ of the real \textit{Gaia} sample. The overlap in discoveries in the simulation is consistent within uncertainties to the real novae.

\begin{figure*}
\begin{center}
 \includegraphics[width=1.0\textwidth]{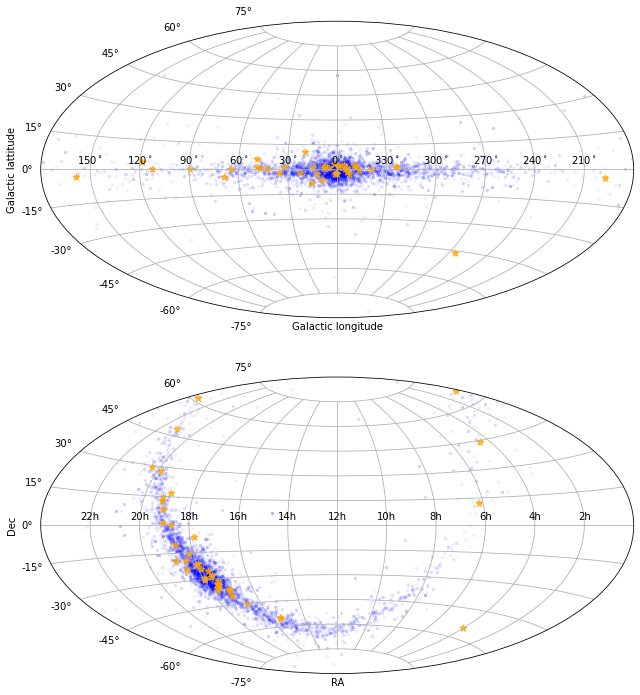}
\caption{The positions of simulated novae that are discovered in our model (blue dots) compared to the real sample discovered between $2019-2021$ (orange stars). The simulated positions are derived assuming novae trace the stellar density of the Galaxy. The elongated bulge, oriented at a roughly 20$^{\circ}$ angle from the Sun--Galactic center line, appears to place more recoverable bulge novae at $l > 0^\circ$ compared to $l < 0^\circ$.}
\label{fig:skymap}
\end{center}
\end{figure*}

The sky positions of the simulated and real samples of novae are shown in Figure~\ref{fig:skymap}. Again, the degree to which the positions agree is hard to assess because of low number statistics, but there is broad agreement between the distribution of simulated and observed novae. Notably, both the observed and simulated nova populations in the bulge region show more novae at Galactic longitude $l > 0^{\circ}$ than at $l < 0^{\circ}$. This is likely due to the elongated bulge (``bar") in this region of the Galaxy, which places typical novae at $l > 0^{\circ}$ at closer distances, and behind less dust, than those at $l < 0^{\circ}$.

The apparent exception to the agreement between model and data is in the southern region of the Galactic plane, where no novae were observed between an RA of 8 and 14 hours from 2019--2021. This could suggest a bias against discovering plane novae in the south compared to the north, which would not be expected given that both ASAS-SN and \emph{Gaia} are all-sky surveys. This discrepancy could also primarily reflect small number statistics, since
in 2017 and 2018, just before the timespan of this survey, several novae in this southern plane region were indeed discovered (V906 Car, V357 Mus, V549 Vel, V1405 Cen), which would tend to equalize the statistics.

\begin{figure*}
\begin{center}
 \includegraphics[width=1.0\textwidth]{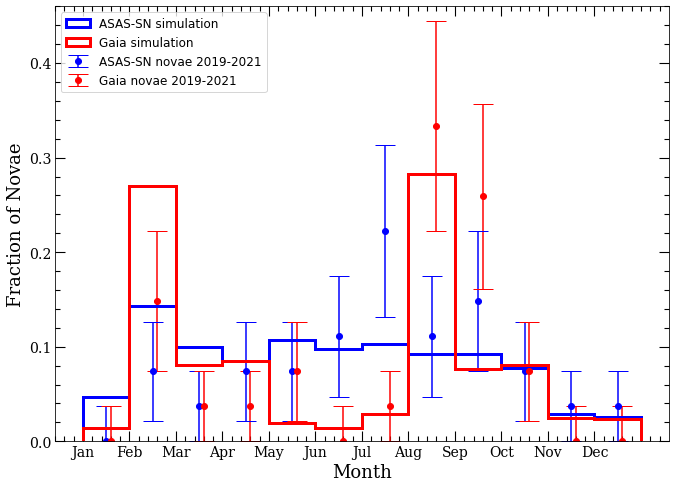}
\caption{Comparison of the month of first detections of model novae (solid histograms) vs. discovered samples (scatter points with Poisson error bars). Both surveys have a seasonal gap while the Sun is in Sagittarius from November to January, and \textit{Gaia} (shown in red) has an additional seasonal gap six months later because of the rotation direction of the satellite. When a field comes out of Solar constraint, the model predicts an excess of nova discoveries.}
\label{fig:months}
\end{center}
\end{figure*}

In addition to comparing where novae are found, we also looked at when novae are discovered by the surveys. Although the simulated novae are given random eruption dates uniformly between 2019 and 2021, we do not expect them to be discovered uniformly throughout the year due to annual changes in observing conditions. When the novae are first detected by ASAS-SN and \textit{Gaia} in our models are shown as histograms in Figure~\ref{fig:months}, along with the observed first detections of novae from Table \ref{table:nova1} with $1\sigma$ Poisson errors. From November to January, the Galactic center region is behind the Sun, and therefore much of the Galaxy is not observable. In February, this field becomes observable again and the novae still bright enough for detection can be discovered, resulting in our model predicting this month to have the highest discovery rate for ASAS-SN and one of the highest for \textit{Gaia}. These are the only pronounced patterns in the predicted annual discovery rate of ASAS-SN. The satellite observations of \textit{Gaia} also have a second area of avoidance around Solar opposition (see Figures 4 and 5 of \citealt{gaia18}). This causes an annual pattern of two \emph{Gaia} nova discovery seasons lasting roughly three months, with over half of first detections happening in February and August when the Galactic center comes out of the areas of avoidance. The ASAS-SN and \emph{Gaia} detections track relatively well with the model predictions but with large uncertainty because of the small number of novae per monthly bin. 

\section{Comparisons to Previous Results}
\label{sec:compare}
Our predicted rate of $26\pm 5$ yr$^{-1}$ is notably lower than most recent direct Galactic nova rate estimates (see Figure~\ref{fig:rate_hist}). It is perhaps consistent with the 
naked eye nova direct extrapolation rate of $50_{-23}^{+31}$ yr$^{-1}$ \citep{shafter_2017}
but mostly due to the latter's large uncertainty. It is inconsistent with the rescaled (because of faint fast novae) M31 inferred Galactic rate of $\sim50$ yr$^{-1}$ to $\sim70$ yr$^{-1}$ performed in the same work.

\subsection{Bulge vs.~Disk Rates}
In our primary model, we assume novae occur proportional to the stellar mass in both the bulge and disk.
Because roughly $40\%$ of the novae are in the bulge region ($R < 3$ kpc) compared to the $60\%$ of novae that would be considered to be in the disk ($R > 3$ kpc), this predicts a bulge and disk rate of ${\rm{bulge}} \approx 10\pm2$ yr$^{-1}$ and ${\rm{disk}} \approx 16\pm2$ yr$^{-1}$. 
There is an independent estimate of the bulge nova rate from the OGLE-IV survey, which is  $13.8 \pm 2.6$ yr$^{-1}$ \citep{mroz15}, which is just over $1\sigma$ higher than our value.

We find little difference ($\lesssim10\%$) in the detection efficiencies between the bulge and disk in our work, with ASAS-SN predicted to recover $\sim 37\%$ of disk novae and $\sim 28\%$ of bulge novae,  and \textit{Gaia} recovering $\sim 43\%$ of disk novae and $\sim 40\%$ of bulge novae. This means that our inferred
total Galactic nova rate would not meaningfully change for reasonable differing assumptions about the bulge and disk populations of novae \citep{kawash21b}.
As an example, \citet{shafter01} estimated that in M31 the nova rate per unit luminosity of the disk was 0.4 that of the bulge. Assuming a similar ratio for the mass in our model yields rates of ${\rm{bulge}} \approx 15$ yr$^{-1}$ and ${\rm{disk}} \approx 11$ yr$^{-1}$. So, even a mild enhancement of the bulge nova rate with respect to the disk nova rate could bring our results into full agreement with the OGLE-IV results, but as stated above the disagreement is small even with our standard model.

We do note that
our inferred rate of disk novae is not consistent even at the $2\sigma$ level with the average disk rate estimate of $67_{-17}^{+21}$ yr$^{-1}$ from \citet{ozdonmez18}. This rate was derived from an estimate of the local (within 1--2 kpc) nova population extrapolated to the entire disk. The origin of the disagreement is not immediately clear, but could potentially arise if the distances to some of the novae in their sample were underestimated.

\subsection{Comparison to Palomar Gattini-IR}
\label{sec:pgir}

The rates derived in this work are perhaps most easily comparable to the PGIR estimate of
$44_{-9}^{+20}$ yr$^{-1}$ \citep{de21}, the only other direct Galactic estimate made from a systematic large (though not all-sky) time domain survey. However, the PGIR rate is substantially higher than those derived here, with the overlap in the posterior distribution functions $< 20\%$. Here, we assess whether the discrepancy can be explained by different assumptions between the two models.

\cite{kawash21b} explored the sensitivity of the ASAS-SN derived nova rates to various model assumptions and found that the rates are most sensitive to the assumed extinction law ($A_V/A_{K_s} = 13.44$ vs. $A_V/A_{K_s} = 8.65$; \citealt{nataf16}) and the stellar density model used to distribute the nova positions (\citealt{robin03} vs. \citealt{catun20}). Because a steeper reddening law increases the $g$-band extinction and the \cite{catun20} model places more novae closer to the plane (see Figure 1 of \citealt{kawash21b}), the recovery fraction of novae in ASAS-SN decreases, resulting in a $33\%$ increase in the derived rate. Assuming a similar increase for this work would increase the ASAS-SN derived Galactic nova rate to $R \approx 37$ yr$^{-1}$, still consistent with our primary model at the 2.2$\sigma$ level and now consistent with the PGIR rate at the 1$\sigma$ level. However, the \textit{Gaia} rate is much less sensitive to these changes as it is better at finding extinguished novae in the plane, so a significant discrepancy between the \textit{Gaia} and PGIR rates would remain.

As seen in Table 1 of \cite{kawash21b}, the PGIR predicted rate is not sensitive to changes in the distribution of positions, extinction model, or bulge-to-disk ratios of novae. Those rates were derived assuming a detection efficiency of $17\%$, estimated from the completeness study of PGIR data \citep{de21}. 

Arguably the largest difference between the PGIR study and this work is the assumed shapes of nova lightcurves and the adherence to the MMRD relation. We assume that the distribution of speed classes is derived from the log-normal distribution of $t_2$ measured in \cite{kawash21}, then use a real nova lightcurve with the closest value of $t_2$ as a template to model the fading from maximum light. \cite{de21} uses the peak luminosity to find the corresponding $t_3$ value using the MMRD relationship measured in \cite{ozdonmez18} and then assumes the apparent magnitude will fade linearly in time. We rerun our analysis, this time using the PGIR method to derive the lightcurve speed and shape, to see if this could explain the discrepancy in the results. If we do this, the derived rates marginally decrease, with the ASAS-SN, \textit{Gaia}, and joint rates changing to $26 \pm 5 ~ \rm{yr}^{-1}$, $21 \pm 5 ~ \rm{yr}^{-1}$, and $24 \pm 4 ~ \rm{yr}^{-1}$, respectively. This minor decrease in the derived rate is because the MMRD relation measured in \cite{ozdonmez18} maps the average nova luminosity of $M = -7.2$ mag to a relatively slow decline t$_3 \approx 73$ days, and the strict adherence to the MMRD relation does not allow for faint and fast novae. This makes novae observable for a slightly longer period of time compared to the method used in this work. But overall, the differences in the shape of the simulated nova lightcurves cannot explain the inconsistencies in the results.

One of the three major requirements for a nova candidate to be identified in the PGIR pipeline is that the transient has to be 3 mags brighter than any 2MASS counterparts within a radius of 10$^{\prime\prime}$ \citep{de21}. From the positions and peak brightness of PGIR-detected novae listed in Table 1 of \cite{de21}, it appears that PGIR\,20ekz and PGIR\,20eig do not meet this requirement. They could have been discovered because the astrometry on individual epochs varies significantly (because of the 8.7$^{\prime\prime}$ PGIR pixel scale), so the positions of a transient from a few observations can meet all of the requirements even when the median position does not. We estimated the degree to which this matters for the recovery fraction in a simulation by studying the 2MASS counterparts near the 10,000 positions of our simulated sample.

To mimic the PGIR detectors, we estimated peak observed brightness of 10,000 simulated novae only using the ASAS-SN site in Texas (similar in latitude to the single Palomar site of PGIR) and ignoring extinction (as the effects of extinction are much milder in PGIR's $J$ band). We estimated the recovery fraction after requiring that the peak brightness be 3 mag brighter than any 2MASS counterpart within 10$^{\prime\prime}$ of any nova. We then compared that result to one where the positions were allowed to vary between epochs by randomly sampling a normal distribution of RA and Dec. 10 times. ASAS-SN and PGIR have similar pixel scales (8$^{\prime\prime}$ vs 8.7$^{\prime\prime}$), so we sampled the positions from a normal distribution with $\sigma = 1.7^{\prime\prime}$, because the 95\% quantile of the astrometric accuracy of ASAS-SN was found to be 3.4$^{\prime\prime}$ (\citealt{ygo19}; no similar analysis has been published for PGIR, to our knowledge). Letting the positions vary on individual epochs yields a detection efficiency roughly $7\%$ higher than the static positions. Assuming a similar increase in the PGIR analysis could increase the recovery fraction to $\sim$24\%, suggesting a derived nova rate of $R \approx 32$ yr$^{-1}$, consistent with this work at the 1.2$\sigma$ level and consistent with the default PGIR rate at the 1.3$\sigma$  level. Another way to look at how sensitive the PGIR rate is to the 2MASS requirement is by decreasing the PGIR sample from 11 to 9 novae (by excluding 20ekz and 20eig). As seen in Figure 8 of \cite{de21}, this would give larger Poissonian uncertainties, corresponding to a $1\sigma$ range of the Galactic nova rate between 27 yr$^{-1}$ and 52 yr$^{-1}$, consistent with both rates within the uncertainties. So, the rates derived in this work are inconsistent at the $1\sigma$ level with the PGIR rate, but small changes in either model would appear to bring the results into agreement at a rate of $\sim30$ per year.

\subsection{Extragalactic Comparisons and Faint/Fast Novae}
Because of uncertainties in galaxy stellar masses and the potential variation of nova rate with stellar population parameters, there are challenges associated with comparing indirectly derived rates from extragalactic nova surveys with direct rates from Galactic surveys. However, extragalactic surveys have historically been more complete in at least some dimensions, so the exercise has been common practice in the literature. The most recent survey of M31 suggested a Milky Way rate of $R = 34_{-12}^{+15}$ \citep{darnley06}, consistent with the rates derived here. Another common practice is to estimate the linear correlation between the nova rate and the log luminosity of a sample of galaxies. \cite{della94} studied five galaxies to infer a Galactic nova rate of 24 yr$^{-1}$,  \cite{shafter2000} studied three galaxies to infer a Galactic nova rate of $27_{-8}^{+10}$ yr$^{-1}$, and \cite{di20} compiled measurements from 14 galaxies to infer a Galactic nova rate of $\sim22$ yr$^{-1}$.
These estimates have large uncertainties, but they are notably all consistent with the most likely rate predicted in this work. 

\cite{shara16} argues that extragalactic nova surveys underestimate nova rates because they neglect to account for faint and fast novae, but the present paper shows that missed faint fast novae are not an important source of uncertainty in the Galactic novae rate (see Section \ref{sec:peak_mag} and Figure \ref{fig:mmrd_sim}) compared to other factors such as the foreground extinction. These simulations are consistent with the lack of discovery in ASAS-SN of a substantial population of faint fast novae after years of daily monitoring. While faint fast novae undoubtedly exist at some level and are a harder to detect, current Galactic data provide no evidence for a large, yet to be discovered population of these sources, and it is plausible that extragalactic studies have overestimated their importance.

\section{Conclusions}
\label{sec:conc}
In this work, we have presented the results of the first direct Galactic nova rate analysis using optical transient surveys with all-sky FOVs. The results predict a much lower rate than recent estimates, with the most likely model, built by combining ASAS-SN and \textit{Gaia} observations, estimating a Galactic nova rate of $R = 26 \pm 5$ yr$^{-1}$. This rate is consistent with the derived rates from ASAS-SN and \textit{Gaia} observations individually. Our analysis suggests that rates above 40 yr$^{-1}$ are unlikely unless: (i) novae have a much lower scale height than predicted from stellar density models (subjecting them to higher dust extinction), (ii) typical extinction is much higher than predicted from existing three-dimensional dust models, or (iii) the reporting efficiencies of ASAS-SN and \textit{Gaia} are much lower than indicated from current evidence.

If the Galactic nova rate is $< 30$ yr$^{-1}$, does that have any broader implications for the Galaxy?  \cite{Izzo2015} detected $^7$Li absorption in V1369 Cen, and suggested that, based off the intensity of the line, novae could explain the overabundance of $^7$Li by assuming a slow nova rate of  24 per year. On the other hand, \cite{molaro2016} argues that only 2 novae per year would be necessary to explain the abundance of $^7$Li in the Galaxy. So, it is possible that even a relatively low nova rate could still explain the Galactic abundances of $^7$Li.

Our derived rate suggests that, over the past 5 years, observers have been discovering about half of the Galaxy's nova eruptions, on average. Between 2019--2021, we find that ASAS-SN and \textit{Gaia} observations alone have recovered $\sim 54\%$ of Galactic novae, and individually, ASAS-SN recovers $\sim 33\%$ of novae and \textit{Gaia} recovers $\sim 42\%$ of novae. 

Though direct estimates of the Galactic nova rate from large time domain surveys allow for fewer assumptions regarding cadence, many assumptions about the pipelines still need to be made. This is likely the reason the rates derived in this work do not agree with the PGIR rate, but we find that small changes in the models can increase the agreement, with the largest overlap occurring around $\sim30$ novae per year.

While it is important to improve our understanding of the efficiency of transient alert pipelines (perhaps though uninformed ``injection" events, as performed by LIGO/Virgo;  \citealt{2012PhRvD..85h2002A}), we can also make progress by gathering additional data from surveys with improved cadences and less sensitivity to dust. As seen in Figure~\ref{fig:skymap}, the number of Northern Hemisphere Galactic plane novae discovered exceeded that in the Southern Plane. Given that there are currently more operating surveys covering the Plane in the north compared to the south, and some northern surveys are less sensitive to dust (such as Palomar Gattini-IR), 
a red or near-IR high-resolution and moderate-cadence survey of the Southern Galactic plane would give close to all-sky coverage of the dusty regions of the Galaxy, allowing the role of dust in nova discovery to be better constrained. Luckily, planned near-IR surveys like the Dynamic REd All-sky Monitoring Survey (DREAMS; \citealt{soon2020}) and the PRime-focus Infrared Microlensing Experiment\footnote{http://www-ir.ess.sci.osaka-u.ac.jp/prime/index.html} (PRIME), along with the multi-band optical Vera Rubin Observatory \citep{lsst}, make the future prospects of southern hemisphere time-domain surveys bright. Depending on the observing strategy to cover the plane, these surveys should have the ability to detect novae not recovered by current observing capabilities. The degree to which these surveys increase the discovery rate will be the next big step in constraining the Galactic nova rate. With additional Southern Hemisphere observations, the transient community could discover up to $\sim80\%$ of the Galactic nova population, leading to a tightly constrained Galactic nova rate.

\section*{Acknowledgements}
AK, LC, EA, and KVS acknowledge financial support of NSF award AST-1751874 and a Cottrell fellowship of the Research Corporation. JS acknowledges support from the Packard Foundation.
BJS, CSK, and KZS are supported by NSF grant AST-1907570. CSK and KZS are supported by NSF grant AST-181440. 
STH is funded by the Science and Technology Facilities Council grant ST/S000623/1.
ZKR acknowledges funding from the Netherlands Research School for Astronomy (NOVA).

We thank the Las Cumbres Observatory and its staff for its continuing support of the ASAS-SN project. ASAS-SN is supported by the Gordon and Betty Moore Foundation through grant GBMF5490 to the Ohio State University, and NSF grants AST-1515927 and AST-1908570. Development of ASAS-SN has been supported by NSF grant AST-0908816, the Mt. Cuba Astronomical Foundation, the Center for Cosmology and AstroParticle Physics at the Ohio State University, the Chinese Academy of Sciences South America Center for Astronomy (CAS- SACA), and the Villum Foundation. 

This work has made use of data from the European Space Agency (ESA) mission \emph{Gaia}\ (\url{https://www.cosmos.esa.int/gaia}), processed by the \emph{Gaia}\ Data Processing and Analysis Consortium (DPAC,
\url{https://www.cosmos.esa.int/web/gaia/dpac/consortium}). Funding for the DPAC has been provided by national institutions, in particular the institutions participating in the \emph{Gaia}\ Multilateral Agreement. Further details of funding authorities and individuals contributing to the success of the mission is shown at \url{https://gea.esac.esa.int/archive/documentation/GEDR3/Miscellaneous/sec_acknowl/}.

This work is in part based on observations obtained at the Southern Astrophysical Research (SOAR) telescope, which is a joint project of the Minist\'{e}rio da Ci\^{e}ncia, Tecnologia e Inova\c{c}\~{o}es (MCTI/LNA) do Brasil, the US National Science Foundation’s NOIRLab, the University of North Carolina at Chapel Hill (UNC), and Michigan State University (MSU).

The analysis for this work was performed primarily in \texttt{ipython} \citep{pg07} using \texttt{numpy} \citep{oliphant2006guide,van2011numpy}, \texttt{Astropy} \citep{astropy:2018}, \texttt{Matplotlib}  \citep{Hunter:2007}, and \texttt{scipy} \citep{vgo20}.

We thank K.\ De for helpful conversations regarding Galactic novae and transient surveys and Jo Bovy for help with the extinction models.

\bibliographystyle{aasjournal}
\bibliography{biblio}



%




\label{lastpage}
\end{document}